\documentclass[12pt,letterpaper,american]{article}
\usepackage[T1]{fontenc}
\usepackage[latin1]{inputenc}
\usepackage{graphicx}

\makeatletter

\providecommand{\tabularnewline}{\\}

\usepackage{a4wide}
\usepackage{setspace}
\usepackage{graphicx}
\usepackage[scanall]{psfrag}

\usepackage{babel}
\makeatother

\begin{document}

\title{Global Potential Energy Minima of (H$_{2}$O)$_{n}$ Clusters on
Graphite: A Comparative Study of the TIP$N$P ($N=3,\:4,\:5$) Family}

\author{B.~S.~Gonz\'alez, J.~Hern\'andez-Rojas, J.~Bret\'on, and J.~M.~Gomez~Llorente%
\thanks{Corresponding author. \emph{E-mail address}: jmgomez@ull.es %
}\\
 Departamento de F\'{\i}sica Fundamental II\\
 Universidad de La Laguna, 38205 Tenerife, Spain}

\maketitle
\begin{abstract}
The water-graphite interaction potential proposed recently (Gonz\'alez
et al.\emph{ J. Phys. Chem. C} \textbf{2007}, \emph{111}, 14862),
the three TIP$N$P ($N=3,\:4,\:5$) water-water interaction models,
and basin-hopping global optimization are used to find the likely
candidates for the global potential energy minima of (H$_{2}$O)$_{n}$
clusters with $n\leq21$ on the (0001)-surface of graphite and to
perform a comparative study of these minima. We show that, except
for the smaller clusters ($n<6$), for which ab-initio results are
available, the three water-water potential models provide mostly inequivalent
conformations. While TIP3P seems to favor monolayer water structures
for $n<18$, TIP4P and TIP5P favor bilayer or volume structures for
$n>6$. These $n$ values determine the threshold of dominance
of the hydrophobic nature of the water-graphite interaction at the
nanoscopic scale for these potential models.
\end{abstract}
\doublespacing

\section{Introduction}

\label{sec1}The interaction between water and graphite has been the
concern of theoretical and experimental studies. A deep understanding
of the features and properties of this interaction is of great interest
in technological applications \cite{lancaster,zaidi,xu}, environmental
sciences \cite{popo}, and astrophysics \cite{draine}, among other
fields. The establishment of either the hydrophilic or hydrophobic
nature of graphite at nanoscopic scales, which is of particular relevance
in those applications, must be based on the knowledge of this interaction. 

In a previous publication \cite{I} (hereafter referred to as I),
we have developed a model for the water-graphite interaction and found
the likely candidates for the global potential energy minima of (H$_{2}$O)$_{n}$
clusters with $n\leq21$ on the (0001)-surface of graphite. Out of
this model, we have obtained a rather hydrophobic water-graphite interaction
at the nanoscopic scale. As a consequence of this property, the water
component of the lowest graphite-(H$_{2}$O)$_{n}$ minima is quite
closely related to low-lying minima of the corresponding (H$_{2}$O)$_{n}$
clusters. In about half of the cases the geometrical substructure
of the water molecules in the graphite-(H$_{2}$O)$_{n}$ global minimum
coincides with that of the corresponding free water cluster. Exceptions
occur when the interaction with graphite induces a change in the geometry
of the water moiety. Our general conclusions were in agreement with
the sparse experimental \cite{chaca,avgul} and theoretical data \cite{xu,werder,karapetian,lin,sudiarta}.
Besides, the structures of these minima for $1<n\leq6$ coincided
with those provided by empirical \cite{karapetian} and ab initio
calculations \cite{xu,lin}. 

In our study, the water-water interaction was described by the TIP4P
intermolecular potential model \cite{jorge}. The related TIP3P potential
\cite{jorge} was also used to model this interaction for $n\leq6$.
The global minimum structures found for these clusters coincided
with those of the TIP4P model. However, the observed dependence of
the structure of the water-graphite global minima on the structure
of the corresponding free water clusters and the known dependence
of the latter on the water-water interaction model for $n>6$ anticipated
a dependence of the structure of these larger water-graphite clusters
on the form chosen to model the water-water interaction. In I, preliminary
results with the TIP3P model confirmed this prediction. In this article,
we will present the concluding results from our analysis of this dependence
by considering also the TIP5P model \cite{mahoney}. As in I, we will
make use of the water-graphite interaction model developed there and
the basin-hopping method to find the likely candidates for the global
potential energy minima of graphite-(H$_{2}$O)$_{n}$ clusters with
$n\leq21$ and the TIP3P and TIP5P water-water interaction models,
and perform a systematic comparison of the cluster structures found
with these and the TIP4P model.

This paper is organized as follows. In Section \ref{sec2} we summarize
the relevant details of the model developed in I for the water-graphite
interaction. In Section \ref{sec3} we present likely candidates for
the cluster global potential energy minima together with their association
and binding energies for both the TIP3P and TIP5P water-water interaction
models. We shall also compare these global minimum structures with those
found in I for the TIP4P model. Finally, Section \ref{sec4} summarizes
our conclusions.

\section{Summary of the Potential Energy Function}

\label{sec2}The closed-shell electronic structure of both graphite
and water makes an empirical approach to the potential energy surface
(PES) for the water-graphite and water-water interactions particularly
attractive. In I, we wrote the potential energy of a graphite-(H$_{2}$O)$_{n}$
cluster as a sum of two contributions \begin{equation}
V=V_{{\rm ww}}+V_{{\rm wg}},\label{eq:1}\end{equation}
where $V_{{\rm ww}}$ is the sum of pairwise water-water interactions,
and $V_{{\rm wg}}$ is the water-graphite term. For the water-water
interaction, the TIP4P model was the primary choice in I; here we
will study the performance of the TIP3P and TIP5P potentials. All
these models describe each water molecule as the same rigid body with
two positive charges on the hydrogen atoms and either a balancing
negative charge at the oxygen atom (TIP3P) or close to the oxygen
atom (TIP4P), or two balancing negative charges close to the oxygen
atom and out of the molecular plane (TIP5P), together with a dispersion-repulsion
center on the oxygen atom. Hence, $V_{{\rm ww}}$ is a sum of pairwise
additive Coulomb and Lennard-Jones terms. We should remind here that
the TIP$N$P are a family of empirical water-water potentials whose
parameters have been appropriately set so as to reproduce some properties
of the liquid water phase at room temperature. Potentials from these
family have been used in the study of homogeneous water clusters \cite{wales0,kabrede,hartke1,james},
water clusters containing metallic cations \cite{briesta,hartke2},
and water-C$_{60}$ clusters \cite{rojas}.

The water-graphite interaction is written as \begin{equation}
V_{{\rm wg}}=V_{{\rm dr}}+V_{{\rm pol}},\label{eq:2}\end{equation}
where $V_{{\rm dr}}$ is a sum of pairwise dispersion-repulsion terms
between the oxygen and the carbon atoms. Each of these terms is expressed
as a Lennard-Jones potential, whose parameters were obtained using
the standard Lorentz-Berthelot combination rules from the corresponding
parameters for the oxygen-oxygen and carbon-carbon interactions in
TIP$N$P water and Steele \cite{steele} graphene-graphene potentials,
respectively. Specifically, we used the values $\varepsilon_{{\rm CO}}=0.385$
kJ/mol and $\sigma_{{\rm CO}}=3.28$\,\AA\, for the TIP3P,
and $\varepsilon_{{\rm CO}}=0.395$ kJ/mol and $\sigma_{{\rm CO}}=3.26$\,\AA\,
for the TIP5P (see \cite{I} for TIP4P parameters),
which are similar to those derived by
Werder \emph{et al.} \cite{werder} to fit the contact angle for a
water droplet on a graphene surface. A simple analytic form for $V_{{\rm dr}}$
can be obtained using Steele summation method \cite{steele,ocasio}
over the graphite periodic structure by writing the interaction of
a dispersion center with a graphite layer as a Fourier series. The
total repulsion-dispersion interaction is obtained as a sum of such
terms over each graphite layer. We have obtained well converged values
by including the continuum contribution from the two upper layers
and the first corrugation from the first layer. 

In Eq.~(\ref{eq:2}), $V_{{\rm pol}}$ includes the energy associated
with the polarization of graphite due to the electric field of all
the water point charges. This many-body interaction, which turns out
to be smaller than $V_{{\rm dr}}$, was evaluated using a continuous
representation of graphite in terms of two contributions,\begin{equation}
V_{{\rm pol}}=V_{\parallel}+V_{\perp},\label{eq:7}\end{equation}
each one associated, respectively, with the response of graphite to
the electric field component parallel and perpendicular to the graphite
surface. For the first one, $V_{\parallel}$, we assumed that graphite
behaves as a classical conductor, which allowed us to make use of
the image charge method to obtain its analytical form. In order to
evaluate $V_{\perp}$, we associated to the graphite surface a constant
surface polarizability density $\alpha_{\perp}$ such that when an
electric field depending on the surface point and perpendicular to
the layer, $E_{\perp}(x,y)$, is applied, an electric dipole density,
$I(x,y)$, is induced on that layer, with $I(x,y)=\alpha_{\perp}E_{\perp}(x,y)$.
Using image charge methods one readily shows that if the graphite
surface coincides with the plane $z=0$, the induced image of an electric
charge $q_{i}$ at the point $(x_{i},y_{i},z_{i})$ is an electric
dipole $p_{i}=-2\pi\alpha_{\perp}q_{i}$ at the point $(x_{i},y_{i},-z_{i})$
and direction parallel to the $z$ axis. This result can be generalized
additively to the case of several electric point charges to obtain
an analytical form for $V_{\perp}$. The value of the polarizability
density $\alpha_{\perp}$ was estimated from $\varepsilon_{\perp}$,
the relative electric permittivity of graphite for applied electric
fields perpendicular to the (0001) surface, whose value is $\varepsilon_{\perp}=5.75$;
namely, \begin{equation}
\alpha_{\perp}=\frac{d(\varepsilon_{\perp}-1)}{4\pi\varepsilon_{\perp}},\label{eq:10}\end{equation}
where $d=3.35$ $\textrm{\AA}$ is the graphite interlayer distance.
We obtained by this procedure $\alpha_{\perp}=0.220$ $\textrm{\AA}$. 

All other electrostatic contributions to the water-graphite interaction
energy having vanishing continuous terms (as the water-charge carbon-quadrupole
interaction) have been neglected, as well as the McLachlan substrate
mediated water-water interaction \cite{bruch}. This potential energy
surface was argued to be superior to previous empirical models \cite{karapetian}.

\section{Global Potential Energy Minima}

\label{sec3}Likely candidates for the global potential energy minima
of graphite-(H$_{2}$O)$_{n}$ clusters with $n\leq21$ were located
using the basin-hopping scheme \cite{wales}, which corresponds to
the `Monte Carlo plus energy minimization approach of Li and Scheraga
\cite{li}. This method has been used successfully for both neutral
\cite{wales} and charged atomic and molecular clusters \cite{rojas,wales2,wales3,rojas1,rojas2},
along with many other applications \cite{wales4}; of course, this was the
method used in I. In the size range considered here the global optimization
problem is relatively straightforward. The global minimum is generally
found in fewer than $7\times10^{4}$ basin-hopping steps, independently
of the random starting geometry. In some cases, starting out from
the (H$_{2}$O)$_{n}$ global potential minimum, the corresponding
global minimum for graphite-(H$_{2}$O)$_{n}$ is found even faster.
However, the success hit rate of the optimization method decreases
significantly and the likelihood of our candidates decreases. As a matter
of fact, we have been able to find for $n=19$ and $n=21$, in each case,
a TIP4P global minimum candidate with energy lower than the one found in I;
although these new candidates present structures very similar
to the previously reported ones.

For graphite-(H$_{2}$O)$_{n}$ clusters, association energies, $\Delta E_{{\rm a}}$,
are defined for the process \begin{equation}
\mbox{graphite}+n\textrm{H}_{2}\textrm{O}=\mbox{graphite-}(\textrm{H}_{2}\textrm{O})_{n};\qquad-\Delta E_{{\rm a}}.\label{r1}\end{equation}
 We also define the water binding energy, $\Delta E_{{\rm b}}$, as
the difference between the association energies of graphite-(H$_{2}$O)$_{n}$
and (H$_{2}$O)$_{n}$; that is, \begin{equation}
\mbox{graphite}+(\textrm{H}_{2}\textrm{O})_{n}=\mbox{graphite-}(\textrm{H}_{2}\textrm{O})_{n};\qquad-\Delta E_{{\rm b}}.\label{r2}\end{equation}
 The clusters in these expressions are assumed to be in their global
minimum. The structures and association energies employed here for
the global minima of (H$_{2}$O)$_{n}$ coincide precisely with those
obtained by Wales and Hodges \cite{wales0}, Kabrede and Hentschke
\cite{kabrede}, and James et al. \cite{james}. %
\begin{figure}
\psfrag{A}[tc][tc]{(a)}
\psfrag{B}[tc][tc]{(b)}
\includegraphics[width=8.25cm]{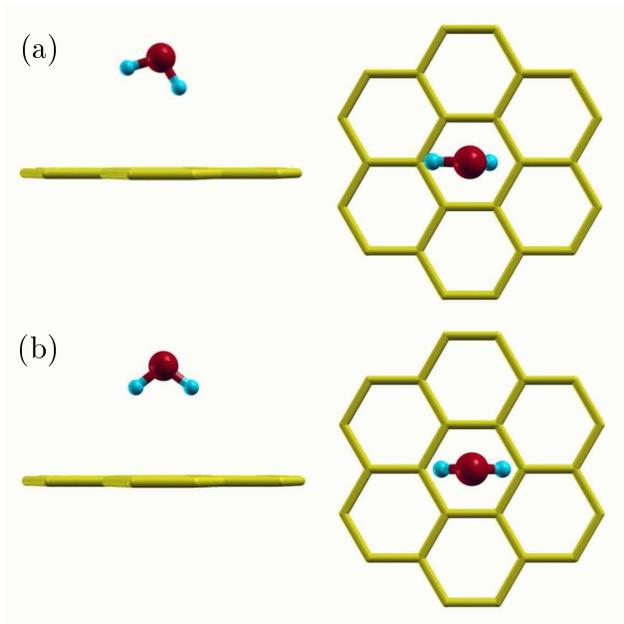}

\caption{\label{f1} Two views of the global minimum obtained for graphite-(H$_{2}$O).
Water-water potential models: TIP3P (a); TIP5P (b). This figure, as well as 
figures \ref{f4} and \ref{f5},
was prepared using the program XCrysDen \cite{xcrys}. }

\end{figure}

In the water monomer case, the structures found for the water-graphite system
with the TIP3P and TIP5P potentials are given in Fig. \ref{f1} (those
for TIP4P were presented in I). While TIP3P, as well as TIP4P \cite{I}, favors
a one-legged structure, the TIP5P model produces a two-legged global
minimum. The equilibrium distance in the global minimum between the
oxygen and the graphite surface is 3.13\,\AA\, for the TIP3P
potential and 3.12\,\AA\, for the other two models; these
distances are very close to the \emph{ab initio} value (3.04\,\AA)
\cite{lin} and the corresponding values in water-C$_{60}$ (3.19\,\AA)
\cite{rojas} and water-benzene (experimental, 3.33\,\AA)
\cite{guto}. As happens with the TIP4P model, one-legged and two-legged
stable structures exist very close in energy for each of the TIP$N$P
models. Therefore, as we have discussed in I, the structures found
here for the water monomer might change by the inclusion in our potential
energy surface of the corrugation terms associated with the electrostatic
interactions, which have been neglected in our PES. For instance,
the carbon-quadrupole contribution may favor a two-legged structure
\cite{steele2}. However, these effects will tend to average out in
the adsorption of water clusters.

The three model potentials provide
very similar monomer binding energies, namely $\Delta E_{{\rm b}}=8.81$\,kJ/mol
for the TIP4P and $\Delta E_{{\rm b}}=8.94$\,kJ/mol for the other two
models, in good agreement with the ab-initio data. The contribution
of the polarization energy to these binding energies ($\sim25\%$)
follows the same trend as the magnitude of the water dipole moment
for each model and it is responsible for the orientation of the H$_{2}$O
molecule on the graphite surface. 

\begin{figure}
\psfrag{Energy}[b][t]{Energy (kJ/mol)}
\psfrag{n}[t][b]{$n$}
\psfrag{1}[t][t]{1}
\psfrag{3}[t][t]{3}
\psfrag{5}[t][t]{5}
\psfrag{7}[t][t]{7}
\psfrag{9}[t][t]{9}
\psfrag{11}[t][t]{11}
\psfrag{13}[t][t]{13}
\psfrag{15}[t][t]{15}
\psfrag{17}[t][t]{17}
\psfrag{19}[t][t]{19}
\psfrag{21}[t][t]{21}
\psfrag{10}[c][l]{}
\psfrag{20}[c][l]{20}
\psfrag{30}[c][l]{}
\psfrag{40}[c][l]{40}
\psfrag{50}[c][l]{}
\psfrag{60}[c][l]{60}
\psfrag{70}[c][l]{}
\psfrag{80}[c][l]{80}
\psfrag{90}[c][l]{}
\psfrag{100}[c][l]{100}
\psfrag{110}[c][l]{}
\psfrag{120}[c][l]{120}
\includegraphics[width=8.25cm]{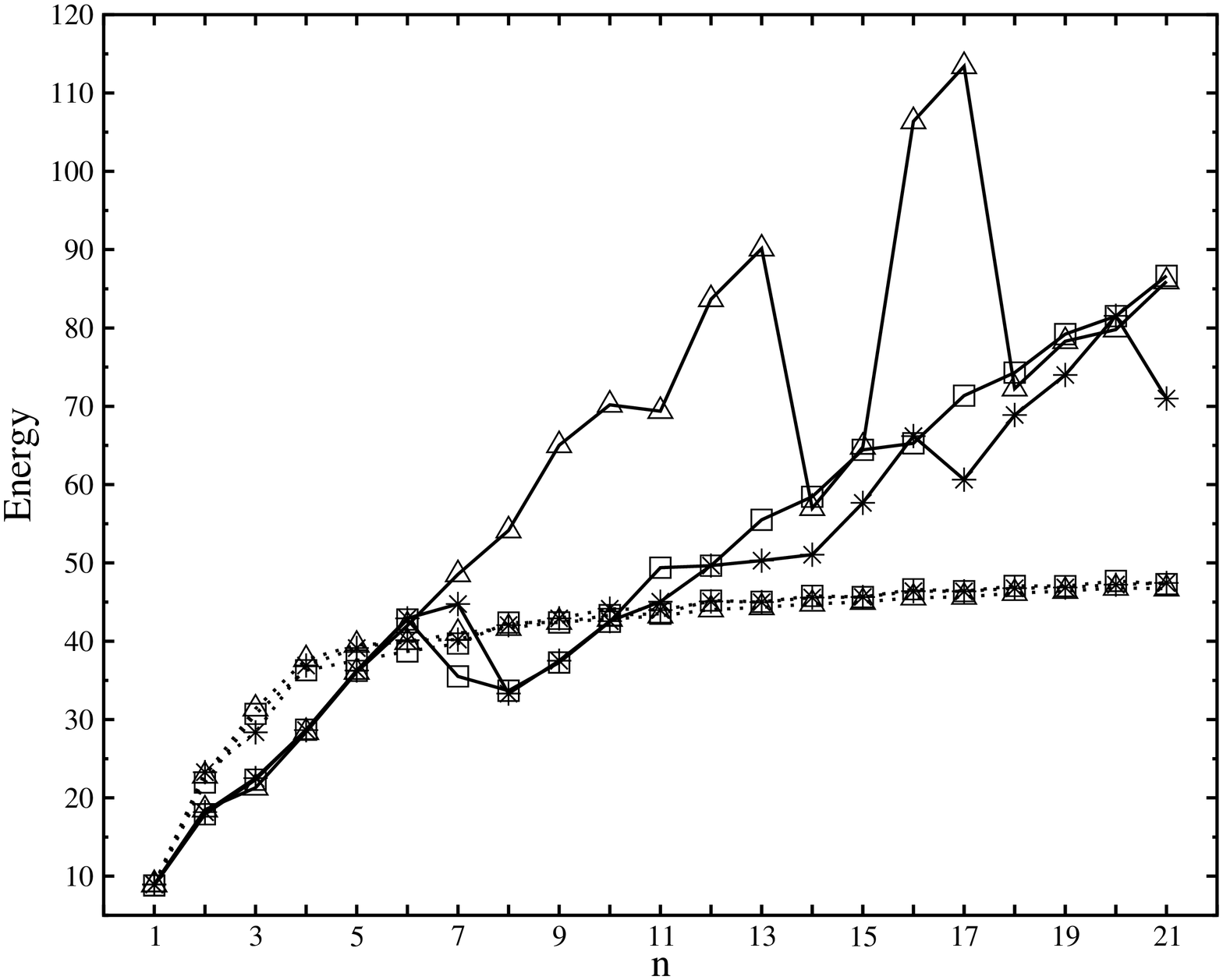}

\caption{\label{f2}Binding, $\Delta E_{{\rm b}}$ (full lines), and association,
$\Delta E_{{\rm a}}/n$ (dotted lines), energies in kJ/mol for the
global minima of water-graphene clusters: TIP3P (triangles), TIP4P (squares),
TIP5P (stars).}

\end{figure}

\begin{figure}
\psfrag{Energy}[b][t]{Energy (kJ/mol)}
\psfrag{N}[t][b]{$n$}
\psfrag{1}[t][t]{1}
\psfrag{3}[t][t]{3}
\psfrag{5}[t][t]{5}
\psfrag{7}[t][t]{7}
\psfrag{9}[t][t]{9}
\psfrag{11}[t][t]{11}
\psfrag{13}[t][t]{13}
\psfrag{15}[t][t]{15}
\psfrag{17}[t][t]{17}
\psfrag{19}[t][t]{19}
\psfrag{21}[t][t]{21}
\psfrag{0}[c][l]{0}
\psfrag{10}[c][l]{}
\psfrag{20}[c][l]{20}
\psfrag{30}[c][l]{}
\psfrag{40}[c][l]{40}
\psfrag{50}[c][l]{}
\psfrag{60}[c][l]{60}
\psfrag{70}[c][l]{}
\psfrag{80}[c][l]{80}
\psfrag{90}[c][l]{}
\psfrag{100}[c][l]{100}
\psfrag{110}[c][l]{}

\includegraphics[width=8.25cm]{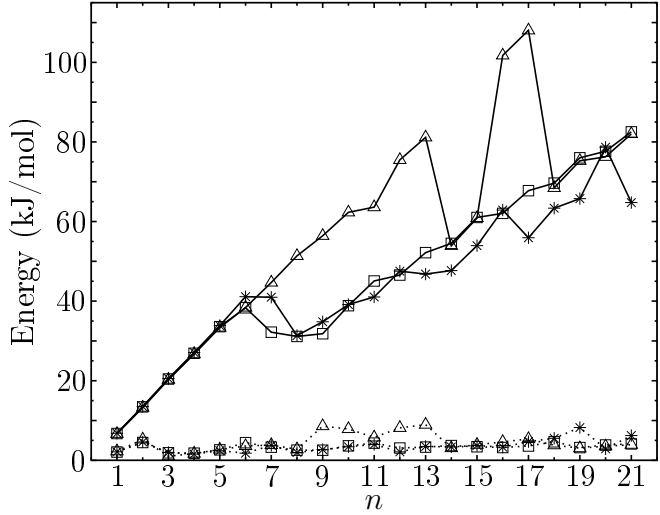}

\caption{\label{f3}Dispersion-repulsion, $V_{{\rm dr}}$ (full line), and
Polarization, $V_{{\rm pol}}$ (dotted line), contributions to the
binding energies in kJ/mol: TIP3P (triangles), TIP4P (squares), TIP5P
(stars).}

\end{figure}

The association ($\Delta E_{{\rm a}}/n$) and binding energies ($\Delta E_{{\rm b}}$)
for the full graphite-(H$_{2}$O)$_{n}$ clusters obtained for the
three TIP$N$P water-water interactions are given in Table \ref{t1}%
\begin{table}
\caption{\label{t1}Global minimum association and binding energies in kJ/mol. }
\begin{tabular}{|c|c|c|c|c|c|c|c|}
\hline 
$n$ & \multicolumn{2}{c|}{TIP3P} & \multicolumn{2}{c|}{TIP4P} & \multicolumn{2}{c|}{TIP5P} & Equivalences\tabularnewline
\cline{2-3} \cline{4-5} \cline{6-7} 
 & $\Delta E_{{\rm a}}$  & $\Delta E_{{\rm b}}$  & $\Delta E_{{\rm a}}$  & $\Delta E_{{\rm b}}$  & $\Delta E_{{\rm a}}$  & $\Delta E_{{\rm b}}$  & \tabularnewline
\hline
\hline 
2 & -45.641 & -18.495 & -43.921 & -17.909 & -46.557 & -18.173 & 3P, 4P, 5P\tabularnewline
3 & -94.128 & -21.292 & -92.182 & -22.317 & -85.108 & -22.521 & 3P, 4P, 5P\tabularnewline
4 & -150.719 & -28.388 & -145.199 & -28.708 & -147.586 & -28.650 & 3P, 4P, 5P\tabularnewline
5 & -197.572 & -36.069 & -187.683 & -36.220 & -195.660 & -36.262 & 3P, 4P, 5P\tabularnewline
6 & -239.280 & -42.001 & -232.207 & -42.773 & -240.824 & -42.918 & 3P, 4P\tabularnewline
7 & -287.249 & -48.535 & -277.938 & -35.512 & -281.399 & -44.749 & 3P, 5P\tabularnewline
8 & -333.402 & -54.132 & -339.035 & -33.686 & -336.466 & -33.311 & 4P, 5P\tabularnewline
9 & -382.078 & -65.013 & -381.527 & -37.277 & -386.252 & -37.510 & 4P, 5P\tabularnewline
10 & -428.329 & -70.183 & -433.315 & -42.486 & -441.510 & -42.560 & 4P, 5P\tabularnewline
11 & -475.783 & -69.380 & -478.621 & -49.388 & -483.825 & -45.044 & \tabularnewline
12 & -528.547 & -83.655 & -542.278 & -49.645 & -540.978 & -49.660 & \tabularnewline
13 & -576.331 & -90.123 & -585.446 & -55.506 & -584.911 & -50.291 & \tabularnewline
14 & -626.761 & -56.986 & -641.126 & -58.448 & -638.105 & -51.048 & 3P, 4P\tabularnewline
15 & -675.072 & -64.806 & -684.664 & -64.439 & -684.710 & -57.668 & 4P, 5P\tabularnewline
16 & -728.942 & -106.378 & -746.178 & -65.272 & -742.269 & -66.191 & \tabularnewline
17 & -776.162 & -113.412 & -788.699 & -71.355 & -788.253 & -60.630 & \tabularnewline
18 & -830.627 & -72.256 & -847.050 & -74.300 & -841.842 & -68.892 & \tabularnewline
19 & -881.578 & -78.300 & -894.134 & -79.740 & -890.437 & -73.994 & 3P, 4P\tabularnewline
20 & -935.993 & -79.786 & -953.950 & -81.513 & -944.519 & -81.551 & 4P, 5P\tabularnewline
21 & -980.995 & -85.926 & -993.960 & -87.051 & -996.230 & -70.984 & 3P, 4P\tabularnewline
\hline
\end{tabular}
\end{table}
 and plotted in Fig. \ref{f2}. %
We also show in Fig.~\ref{f3} the values of the polarization energy
$V_{{\rm pol}}$ and water-graphite dispersion-repulsion energy $V_{{\rm dr}}$,
as defined in Section \ref{sec2}, for the cluster global minima.
The three water-water potential models provide very similar values
for the association energies. However, the binding energies for TIP4P
and TIP5P models differ significantly from those obtained with the
TIP3P for $n>6$. The origin of this difference is, as can be seen
in Fig.~\ref{f3}, in the $V_{{\rm dr}}$ term, which is the dominant
contribution for $n>2$. The term $V_{{\rm pol}}$ oscillates with
$n$ around an average value of $\overline{V}_{{\rm pol}}=3.5$\,kJ/mol;
the two contributions to $V_{{\rm pol}}$, $V_{\parallel}$ and $V_{\perp}$,
are similar in magnitude with $V_{\parallel}$ somewhat larger than
$V_{\perp}$. The term $V_{{\rm dr}}$ fluctuates also around a slowly
growing average as the number of water molecules close to the graphite
surface increases.%
 On average, each of these water molecules contributes about $7.3$\,kJ/mol
to $V_{{\rm dr}}$. The water-graphite binding energies correspond
quite closely to the sum of $V_{{\rm pol}}$ and $V_{{\rm dr}}$,
while the association energies are dominated by the water-water interaction.
The average value of the association energy per molecule in homogeneous
TIP$N$P $(\textrm{H}_{2}\textrm{O})_{n}$ clusters with $6\leq n\leq21$
is $\sim42$\,kJ/mol \cite{wales0,kabrede}. For water cluster on
graphite the corresponding value turns out to be $44.6$\,kJ/mol,
which is comparable with the experimental value of $43.4\pm2.9$\,kJ/mol
\cite{chaca}. Any of these values corresponds to the binding energy
of a water molecule in a water cluster, and it is much larger than
the energy for binding a water molecule onto the graphite surface.
This energy balance would support an hydrophobic nature of the water-graphite
interaction at large scale, as we have already discussed in I.

\begin{figure}
\psfrag{2}[][]{$n=2$}
\psfrag{3}[][]{$n=3$}
\psfrag{4}[][]{$n=4$}
\psfrag{5}[][]{$n=5$}
\psfrag{6}[][]{$n=6$}
\psfrag{7}[][]{$n=7$}
\psfrag{8}[][]{$n=8$}
\psfrag{9}[][]{$n=9$}
\psfrag{10}[][]{$n=10$}
\psfrag{11}[][]{$n=11$}
\psfrag{12}[][]{$n=12$}
\psfrag{13}[][]{$n=13$}
\psfrag{14}[][]{$n=14$}
\psfrag{15}[][]{$n=15$}
\psfrag{16}[][]{$n=16$}
\psfrag{17}[][]{$n=17$}
\psfrag{18}[][]{$n=18$}
\psfrag{19}[][]{$n=19$}
\psfrag{20}[][]{$n=20$}
\psfrag{21}[][]{$n=21$}
\includegraphics[height=20cm]{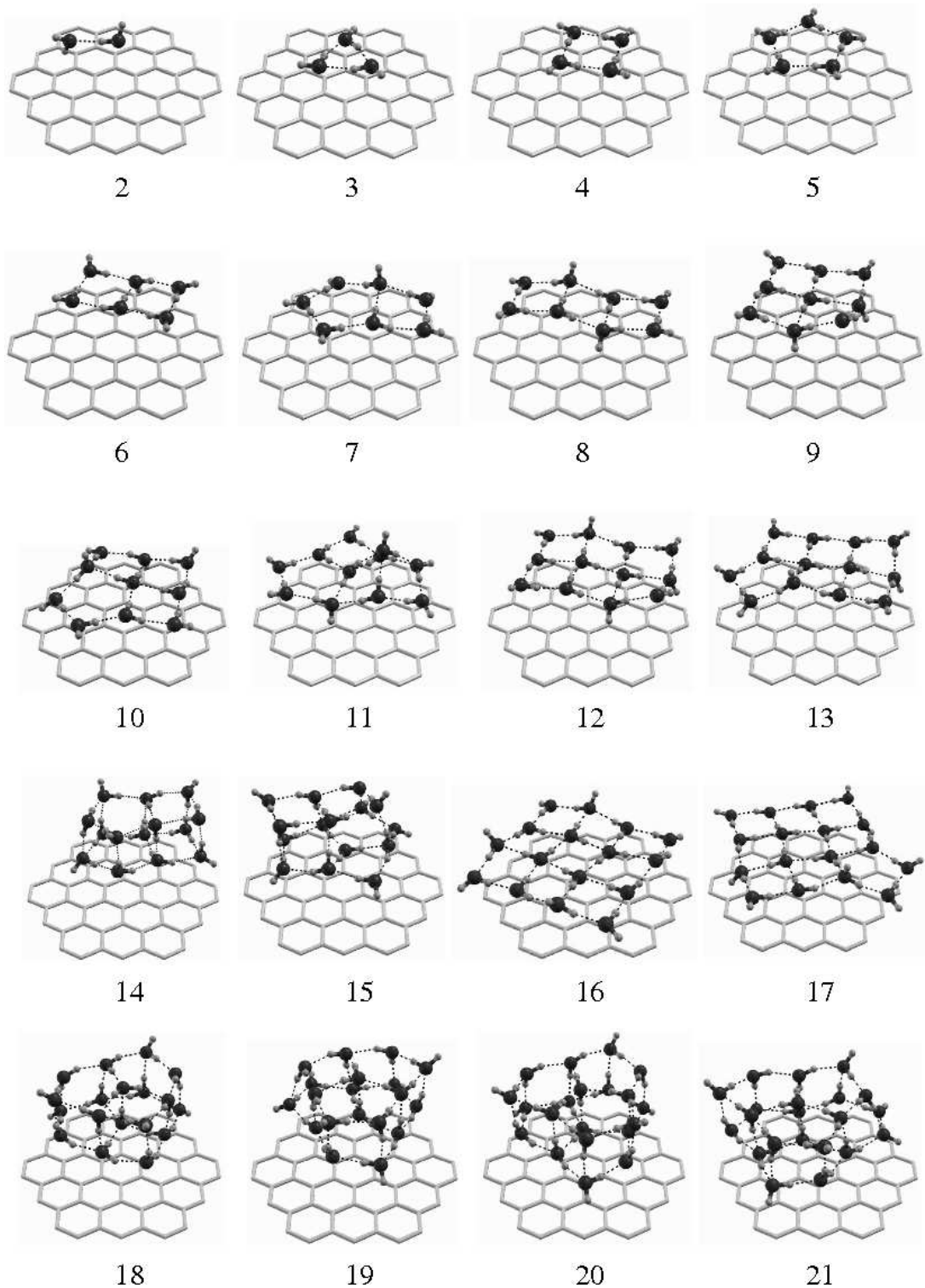}

\caption{\label{f4} Likely global minima obtained for graphite-(H$_{2}$O)$_{n}$
clusters with the TIP3P water-water potential model.}

\end{figure}

\begin{figure}
\psfrag{2}[][]{$n=2$}
\psfrag{3}[][]{$n=3$}
\psfrag{4}[][]{$n=4$}
\psfrag{5}[][]{$n=5$}
\psfrag{6}[][]{$n=6$}
\psfrag{7}[][]{$n=7$}
\psfrag{8}[][]{$n=8$}
\psfrag{9}[][]{$n=9$}
\psfrag{10}[][]{$n=10$}
\psfrag{11}[][]{$n=11$}
\psfrag{12}[][]{$n=12$}
\psfrag{13}[][]{$n=13$}
\psfrag{14}[][]{$n=14$}
\psfrag{15}[][]{$n=15$}
\psfrag{16}[][]{$n=16$}
\psfrag{17}[][]{$n=17$}
\psfrag{18}[][]{$n=18$}
\psfrag{19}[][]{$n=19$}
\psfrag{20}[][]{$n=20$}
\psfrag{21}[][]{$n=21$}
\includegraphics[height=20cm]{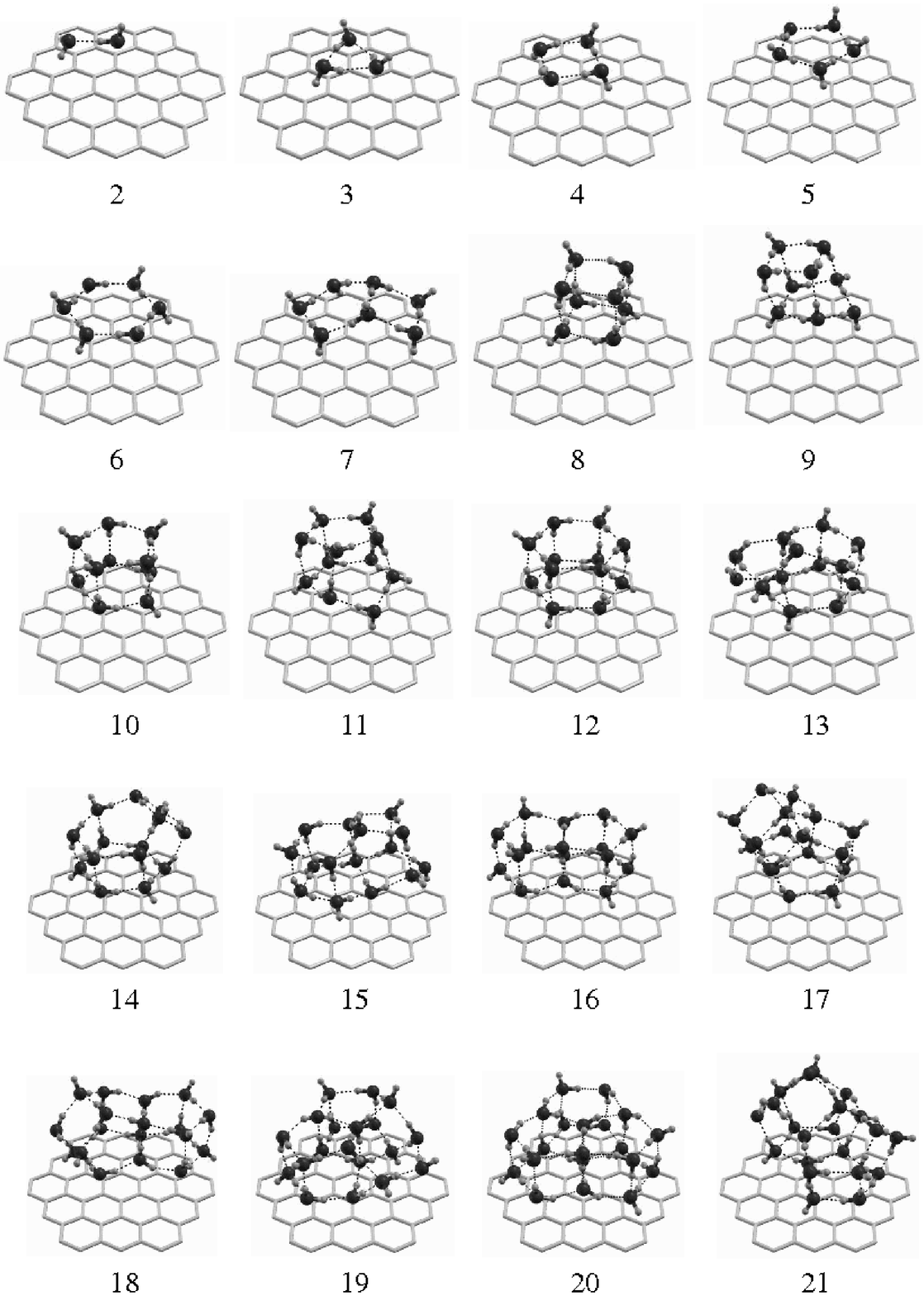}

\caption{\label{f5} Likely global minima obtained for graphite-(H$_{2}$O)$_{n}$
clusters with the TIP5P water-water potential model. }

\end{figure}

The structures of the TIP3P lowest minima obtained for graphite-(H$_{2}$O)$_{n}$
are presented in Fig. \ref{f4}, and those of the TIP5P in Fig. \ref{f5}
(The corresponding TIP4P global minima were presented in I). The three
water-water model potentials provide practically identical structures
for $2\le n\le5$. The water substructures in these compounds are
actually equivalent (see below)
to those in the corresponding free global minimum of TIP$N$P (H$_{2}$O)$_{n}$
\cite{wales0,kabrede,james}, and are in agreement with the ab-initio
results. For $n=6$, TIP3P and TIP4P model have a ``book'' global
minimum, as the one predicted by ab-initio calculations \cite{xu,lin},
while the TIP5P leads to an hexagonal ring. This result would favor
the first two models over the last one. Only the water substructure
in the TIP4P differs from that of the corresponding free water cluster
global minimum (``cage'' conformation). %

Here, we will consider two global minima equivalent
if their water moieties share the same geometrical structure
(aside from minor differences in angles and distances) and orientation
with respect to the graphite surface plane
With this convention the equivalences found in the
global minima of graphite-(H$_{2}$O)$_{n}$ compounds among the three
PES have been included in the last column of Table \ref{t1}. The
number of these equivalences is similar to the number of equivalences
between the three model potentials in the corresponding free water
clusters, although these equivalences involve different clusters and
potentials. For instance, TIP3P and TIP5P provide equivalent free
water global minima for $n=8$; both models present the $D_{2d}$
``cube'' conformation, while TIP4P global minimum has an
$S_{4}$ ``cube'' conformation. On the other hand, the corresponding water
clusters on graphite keep the free structure for TIP4P and TIP5P, while
TIP3P provides a monolayer water conformation; therefore no equivalent
structures appear in this case.

For graphite-water clusters with $6<n\le17$, the TIP3P model seems
to favor monolayer water structures (with exceptions for $n=14$ and
$15$), while the other two models favor either bilayer (TIP4P and TIP5P)
or volume (only TIP5P) structures. The exceptions to the monolayer
pattern for $n=14,15$ in TIP3P, may be explained by the relative
higher stability (deduced from second energy differences) of the corresponding
free water clusters respect to that of their neighbors $n=13,16$.
For $17<n\leq21$, either volume or bilayer conformations are found
for the three model potentials. 

When we compare the conformation of the water substructure on the
graphite surface with that of the corresponding free water cluster,
we find also a markedly different behavior for the TIP3P model. With
the exception of the first six clusters and the case $n=14$, those
two conformations are inequivalent. In other word, the water-graphite
interaction is able to strongly modify the structure of the free water
clusters. This together with the monolayer conformation of the adsorbed
water clusters would point out to a hydrophilic water-graphite interaction
for this potential model. However this behavior seem to be due to
finite size effects since we have not found monolayer global minimum
structures for $n>17$, neither local minima monolayer conformations
that are close in energy to the global minima. For $n>6$ in TIP4P
and $n>7$ in TIP5P, we do not find either such monolayer structures
for these two potential models. Thus the hydrophobic nature of the
water-graphite interaction appears earlier in these models. Furthermore,
in some cases for the two latter models,
the conformation of the water substructure on graphite
and that of the corresponding free water cluster are equivalent. The
exceptions are $n=6,\,7,\,11,\,15,\,17,\,19,\,21$ for TIP4P and $n=7$, $n\geq13$
for TIP5P. In these cases, the water substructure is equivalent to
a low-lying local minimum of the corresponding TIP$N$P (H$_{2}$O)$_{n}$
cluster, rather than to the global minimum. The energy penalty for this
choice is mainly compensated by a more favorable dispersion-repulsion
contribution to the interaction energy with graphite, which arises
from a larger water-graphite contact area. In the structures for the
three model potentials one finds square and pentagonal water rings;
on the other hand, hexagonal rings are less common but they appear
more often in TIP5P ($n=11,\,12,\,14,\,15,\,19$ ) than in TIP3P and
TIP4P (just for $n=21$). 

For TIP4P, the complete two-layer water structures for even $n$ are
precisely the structures of the global TIP4P free water clusters.
Therefore, these structures interact with graphite in an optimal way
and they keep their structure in the corresponding water-graphite
clusters. On the other hand, for odd $n$, the free water global minima
do not show optimal surfaces for its interaction with graphite, thus
explaining why these clusters change their structure to minimize that
interaction energy. The chosen new structures are sensibly determined
by those of either the $n-1$ or $n+1$ clusters. The TIP5P (and also
TIP3P) potential model do not produce this alternating behavior in
the structure of the free water global minima and, therefore, we find
a different behavior in the water-graphite global minima for $n\geq8$.

\begin{figure}
\psfrag{Energy}[b][t]{Energy (kJ/mol)}
\psfrag{\(a\)}[][]{(a)}
\psfrag{\(b\)}[][]{(b)}
\psfrag{n}[t][b]{$n$}
\psfrag{2 }[t][t]{2}
\psfrag{4 }[t][t]{4}
\psfrag{6}[t][t]{6}
\psfrag{8}[t][t]{8}
\psfrag{10}[t][t]{10}
\psfrag{12}[t][t]{12}
\psfrag{14}[t][t]{14}
\psfrag{16}[t][t]{16}
\psfrag{18}[t][t]{18}
\psfrag{20}[t][t]{20}
\psfrag{-3}[c][l]{-3}
\psfrag{-2}[c][l]{-2}
\psfrag{-1}[c][l]{-1}
\psfrag{0}[c][l]{0}
\psfrag{1}[c][l]{1}
\psfrag{2}[c][l]{2}
\psfrag{3}[c][l]{3}
\psfrag{-8}[c][l]{-8}
\psfrag{-6}[c][l]{-6}
\psfrag{-4}[c][l]{-4}
\psfrag{-2}[c][l]{-2}
\psfrag{0}[c][l]{0}
\psfrag{2}[c][l]{2}
\psfrag{4}[c][l]{4}
\includegraphics[width=8.25cm]{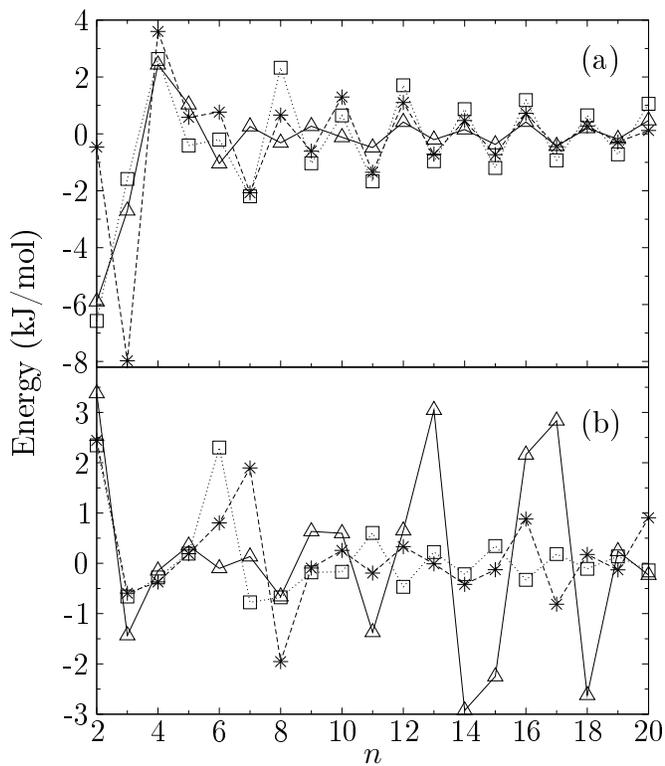}

\caption{\label{f6} Second energy differences per water molecule (in kJ/mol)
for the association energies (a) and binding energies (b) of water-graphite
clusters. Water-water potential models: TIP3P (triangles, full line),
TIP4P (squares, dotted line), TIP5P (stars, dashed line)}

\end{figure}

Second energy differences account for the relative cluster stability;
their values for association and binding energies, per water molecule,
are plotted in Fig. \ref{f6}. %
TIP4P and TIP5P, show practically the same behavior in the whole $n$
range. The $n=4$ cluster is particularly stable in all cases. For
$n>10$, we observe that the three model potentials present an oscillation
of period $\Delta n=2$, namely, clusters with even $n$ are more
stable than their odd $n$ neighbors. This is an interesting feature
because it does not occur so neatly for the free water clusters, and
it is not obviously related, except in the TIP4P case \cite{I}, with the cluster
structures. Second differences for the binding energies do not show
common patterns for the three potential models because these provide
very different global minima structures respect to the water-graphite
interaction.

\section{Conclusions}

\label{sec4}Using basin-hopping global optimization and a potential
energy surface built up from three different water-water interaction
models (TIP3P, TIP4P and TIP5P) we have characterized the geometrical
structures and energetics of the likely candidates for the global potential
energy minima of graphite-(H$_{2}$O)$_{n}$ clusters up to $n=21$.
The structures of these minima for $1<n\leq5$ coincide for the three
potential models with those provided by other available calculations.
The global minimum for the compound with $n=6$ agrees with the ab-initio
structure for TIP3P and TIP4P, but not for TIP5P. For $n>6$, no ab-initio
data are available and, except for the equivalences presented in Table
I, the three model potentials provide different global minimum structures,
as occurs for the free water clusters. For $n>2$, association energies
are dominated by the water-water interaction while the main contribution
to the binding energies comes from the dispersion energy; furthermore
the polarization term $V_{{\rm pol}}$ can be safely neglected for the larger
clusters ($n>3$); this justifies the use of water-graphite potentials
that include only dispersion-repulsion terms \cite{werder}. For small
$n$, the water grows on the graphite surface forming a monolayer.
However, as $n$ increases the hydrophobic nature of the water-graphite
interaction dominates and breaks this tendency. The threshold for
this transition is at $n=7$ for TIP4P and TIP5P and $n=18$ (with
the exceptions $n=14,\,15$) for TIP3P. Therefore this latter potential
seem to favor planar conformations up to larger $n$. 

The hydrophobic character of the water-graphite interaction at the
nanoscopic level makes in some cases the water substructure in the
lowest energy clusters to be equivalent to a low-lying minimum of
the appropriate (H$_{2}$O)$_{n}$ free cluster. In many cases the
structure is simply a slightly relaxed version of the global minimum
for (H$_{2}$O)$_{n}$, and therefore equivalent to it. For TIP3P
this occurs only for the first six $n$ values and for $n=14$. TIP5P
shows equivalences for the same first six $n$ values and for $n=8,\:9,\:10,\:11,\:12$.
TIP4P shows the larger number of equivalences for $n\leq5$ and $n=8,\,9,\,10,\,12,\,13,\,14,\,16,\,18,\,20$.

The lowest energy structures obtained in the present work will be
made available for download from the Cambridge Cluster Database \cite{data}.

\section*{Acknowledgments}

This work was supported by `Ministerio de Educaci\'on y Ciencia (Spain)'
and `FEDER fund (EU)' under contract No. FIS2005-02886. One of us
(BSG) also aknowledges `Ministerio de Educaci\'on y Ciencia (Spain)'
for an FPU fellowship .

\end{document}